\documentclass[10pt, conference, letterpaper]{IEEEtran}
\usepackage[
top    = 0.7in,
bottom = 1in,
left   = 0.68in,
right  = 0.68in]{geometry}
\usepackage{booktabs} % For formal tables

\usepackage{epsfig}
\usepackage{amsmath}
\usepackage{subfigure}
\usepackage{graphicx}
\usepackage{cite}
\usepackage{amssymb}
\usepackage{fixmath}
\usepackage{caption}
\usepackage{algorithmic}
\usepackage{algorithm}
\usepackage[usenames]{color}
\usepackage{dsfont}
\usepackage[super]{nth}
\usepackage{color,soul}

\usepackage{algorithmic}
\usepackage{algorithm}
\usepackage{multirow}
\usepackage{comment}
\usepackage{flushend}
\usepackage{balance}
\usepackage{paralist}

\usepackage{dblfloatfix}
\usepackage{siunitx}

\newdimen\figrasterwd
\figrasterwd\textwidth

\usepackage[normalem]{ulem}

\usepackage{ulem}

\begin{document}
% \title{Development of an Open-source Digital Twin for the POWDER Platform: Methodology and Challenges}

% \title{On the Effects of Model Fidelity on Sim-to-Real Transfer Gap in Twinning the POWDER Platform}

% \title{On the Effects of Modeling on Sim-to-Real Transfer Gap in Twinning the POWDER Platform}

\title{On the Effects of Modeling on the Sim-to-Real Transfer Gap in Twinning the POWDER Platform}

% \title{The Effects of Digital Twin to Real World Gap on Learning Policy Robustness over POWDER Platform}

\IEEEoverridecommandlockouts
\author{Maxwell McManus$^1$, Yuqing Cui$^{1,2}$, Josh (Zhaoxi) Zhang$^1$,\\  Elizabeth Serena Bentley$^2$, Michael Medley$^3$, Nicholas Mastronarde$^1$, Zhangyu Guan$^1$\\
% Dept. of Electrical Engineering, 
% $^1$Department of Electrical Engineering, University at Buffalo, Buffalo, NY 14260, USA\\
% % $^2$Georgia Tech Research Institute (GTRI), Atlanta, GA 30332, USA\\
% $^2$U.S. Air Force Research Laboratory (AFRL), Rome, NY 13441\\
% $^3$SUNY Polytechnic Institute, Utica, NY 13502, USA\\
$^1$Dept. EE, University at Buffalo, USA; $^2$ U.S. AFRL; $^3$SUNY Polytechnic Institute, USA\\
Email: \{memcmanu, yuqingcu, zhaoxizh, nmastron, guan\}@buffalo.edu\\
elizabeth.bentley.3@us.af.mil, michael.medley@sunypoly.edu \vspace{-4mm}
\thanks{This work was supported in part by the National Science Foundation (NSF) under Grant SWIFT-2229563 and the U.S. Air Force Research Laboratory under Contracts FA8750-21-F-1012 and FA8750-20-C-1021.}
\thanks{Distribution A. Approved for public release: Distribution Unlimited: AFRL-2024-4719 on 26 Aug 2024.}
}

\maketitle

\begin{abstract} 
% The focus of this paper will shift from channel modeling to the development process and evaluation of a virtualization of the POWDER platform, investigating how the simulation design methodology and environment virtualization contribute to the observed sim-to-real gap. The results will include analysis of the sim-to-real gap discovered while evaluating the virtual model, as well as proposed solutions to the encountered challenges. 

% \todo{Overall: need to revise the language to emphasize our design process and connections to system identification; simply stating what we have done is not very convincing, emphasize the amount of work done to accomplish this. Same for abstract and full paper body.}
% Digital twin (DT) is envisioned as a key technique for NextG wireless systems, with a great potential of 
% % systems for the wireless domain offer significant promise for NextG networking capabilities, including integration 
% enabling state-of-the-art data-driven control 
% % algorithms 
% and improving dynamic spectrum access capabilities.
Digital Twin (DT) technology is expected to play a pivotal role in NextG wireless systems. 
% , with significant potential to enhance data-driven control and improve dynamic spectrum access capabilities.
% schemes
% among others.
% However, 
% % as of today,
% % an open-source, open access DT system that can support advanced wireless 
% % research on a large scale is still missing. 
% % Furthermore, 
% the evaluation of data-driven algorithms for use in DT, especially transfer learning from simulation to reality, remains an open challenge. 
However, a key challenge remains in the evaluation of data-driven algorithms within DTs, particularly the transfer of learning from simulations to real-world environments. 
% remains an open challenge.
% in this area. 
% \todo{Motivate this work}
% Towards this goal, 
% To fill this gap, 
% In this work we 
% % demonstrate 
% investigate the sim-to-real gap in twinning the NSF PAWR Platform POWDER. 
% % , using an open-source network simulator and several standard analytical models. 
% %  
% We first leverage geographical measurements of the POWDER deployment area to design a 3D model of the University of Utah campus, including all rooftop POWDER nodes. 
% % , then develop a specialized path loss model based on data collected from POWDER to provide realistic simulation performance. 
% We then analyze the accuracy of several path loss models in training data-driven modeling and control policies, and evaluate the sim-to-real gap induced by each model at predicting link performance.
% %
% This analysis is intended to inform the model selection and simulation design process associated with DT-enabled wireless networks, and lay the groundwork for designing a DT for the POWDER platform based on our own experiences and lessons learned in this work. 
% \todo{Abstract will be updated after the main body is finished}
% We also share insight regarding recent use of the POWDER platform and lessons learned for future work enabling DT technology in the wireless domain. 
In this work, we investigate the sim-to-real gap in developing a digital twin for the NSF PAWR Platform, POWDER. We first develop a 3D model of the University of Utah campus, incorporating geographical measurements and all rooftop POWDER nodes. We then assess the accuracy of various path loss models used in training modeling and control policies, examining the impact of each model on sim-to-real link performance predictions. Finally, we discuss the lessons learned from model selection and simulation design, offering guidance for the implementation of DT-enabled wireless networks. 
% This analysis is intended to 
% This work can help guide model selection and simulation design for DT-enabled wireless networks and lays the foundation for the design of a DT for the POWDER platform, informed by the experiences and insights gained in this work.
\end{abstract}


\section{Introduction} \label{sec:intro}

The concept of Digital Twin (DT) has attracted substantial attention as an enabling technology next-generation (NextG) wireless communication systems \cite{mcmanus2023survey}. Recent studies have demonstrated that integrating high-fidelity virtual wireless network models with their configurable real-world counterparts via DT can significantly enhance the performance of NextG networks. This improvement spans a wide range of applications, including computational offloading in mobile edge networks \cite{WenSunReducing20}, UAV network  configuration \cite{LeiLei21uav}, and network resource management \cite{Haozhe20}, among others.

However, the generalizations inherent in network simulations for DT often result in synthetic data that differs significantly from data collected over-the-air (OTA). This discrepancy 
% arises primarily due to the difficulty in accurately simulating the nonlinearities introduced by RF devices. These errors 
further contributes to the sim-to-real gap, which refers to the performance deviation between simulated applications and their real-world deployment in OTA networking scenarios. For data-driven applications, an incomplete understanding of the sim-to-real gap can undermine model accuracy and lead to inefficient tuning of both model and simulation parameters, delaying the development of control policies. 

In this work, using the POWDER platform as a case study, we take an initial step toward rigorous model selection evaluation within the DT design process for wireless networks. Specifically, we analyze the sim-to-real gap of a virtual model of the POWDER platform, generated using an open-source wireless network simulator, by examining two data-driven approaches: predictive modeling and network protocol optimization. We demonstrate the impact of model selection on the performance gap by comparing data-driven policies trained on synthetic data with those trained on real-world data collected from the POWDER platform. 
% Our direct assessment of the sim-to-real gap in DT-enabled wireless networks aims to accelerate the practical adoption of data-driven wireless capabilities in NextG networks.

The primary contributions of this work are as follows: 
\begin{itemize}
    % \item We detail the POWDER testbed virtualization process as an example for future DT design. The virtual testbed includes the full POWDER networking environment, selected hardware, and geographic details, and is implemented in
    % the \underline{U}niversal \underline{B}roadband \underline{Sim}ulator (UBSim),
    % an extensible open-source wireless network simulator. 
\item We outline the virtualization process of the POWDER testbed as a reference for future DT design. The virtual testbed replicates the complete POWDER networking environment, including selected hardware and geographic specifics, and is implemented within the \underline{U}niversal \underline{B}roadband \underline{Sim}ulator (UBSim), an extensible open-source wireless network simulator \cite{NEXTCOMCOM}.

    % \item We use both physical and virtual POWDER testbeds to derive several data-driven policies for two network control problems: first, predicting point-to-point link performance; and second, maximizing link quality-of-service (QoS) via protocol optimization. 
    % We then demonstrate the sim-to-real gap on the resulting policies by comparing the synthetically-trained models with models trained using data collected OTA. 
    \item We leverage both physical and virtual POWDER testbeds to develop data-driven policies for two network control problems: predicting point-to-point link performance and optimizing protocols to maximize link quality-of-service (QoS). We then quantify the sim-to-real gap by comparing policies trained on synthetic data with those trained on over-the-air (OTA) data, and analyze the resulting performance differences between these two approaches.
    % \item We then introduce a portion of OTA data to the synthetic datasets and repeat our analysis to observe the effects of hybrid data collection on the sim-to-real gap. \todo{Can only include if enough time}.
    % \item We discuss the implications of a DT framework for use with the POWDER platform, and make several recommendations for future
    % % research on applications of DT-enabled networking and 
    % use of the POWDER platform for DT-enabled networking based on our experimental findings. 
    \item Based on our experimental findings, we explore the implications of applying a DT framework to the POWDER platform and provide recommendations for leveraging the platform in DT-enabled network design and control. 
    \end{itemize}
The remainder of the paper is organized as follows. 
Section~\ref{sec:related} provides an overview of related works. 
% Section~\ref{sec:powder} presents an overview of the POWDER platform, highlighting key elements relevant to the DT implementation.
% An overview of the POWDER platform is provided and several key elements are highlighted for use in the DT in Section~\ref{sec:powder}. 
% We discuss the DT design process, including data collection and virtual model deployment 
In Section~\ref{sec:design}, we detail the DT design process, including data collection and virtual model deployment.
Section~\ref{sec:datadriven} outlines the experimental approach used to benchmark the proposed DT system and Section~\ref{sec:evaluation} discusses the results of the experimental campaign. 
In Section~\ref{sec:discussion}, we share the lessons learned from the virtualization process. 
% , we outline the experimental approach used to benchmark the proposed DT system, as well as discuss the results of this experimental campaign. 
Finally, we draw the main conclusions in Section~\ref{sec:conclusion}.

\begin{figure*}[b]
\begin{center}
\vspace{-3mm}
\begin{tabular}{cccc}
\frame{\includegraphics[width=0.2\textwidth]{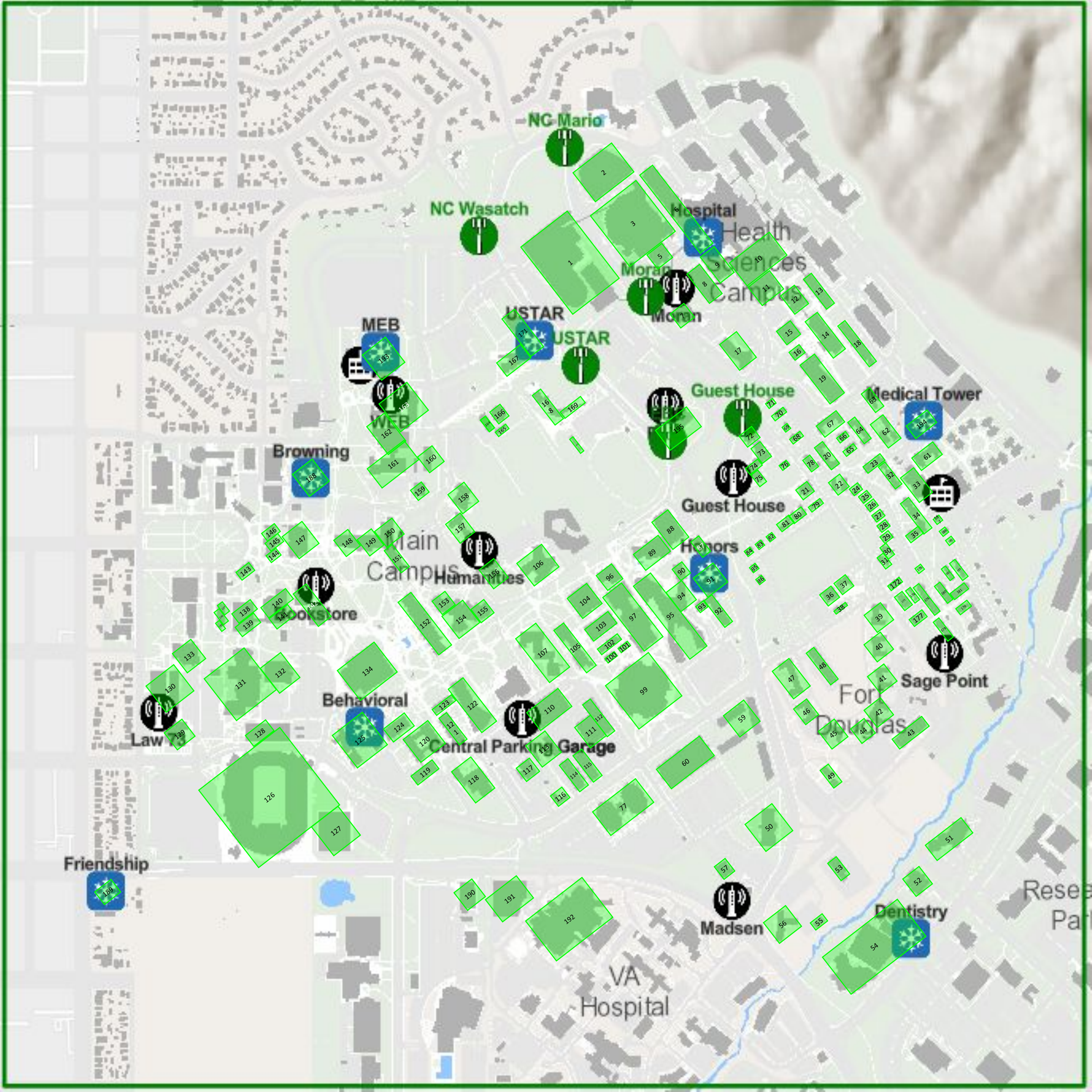}} &
\frame{\includegraphics[width=0.2\textwidth]{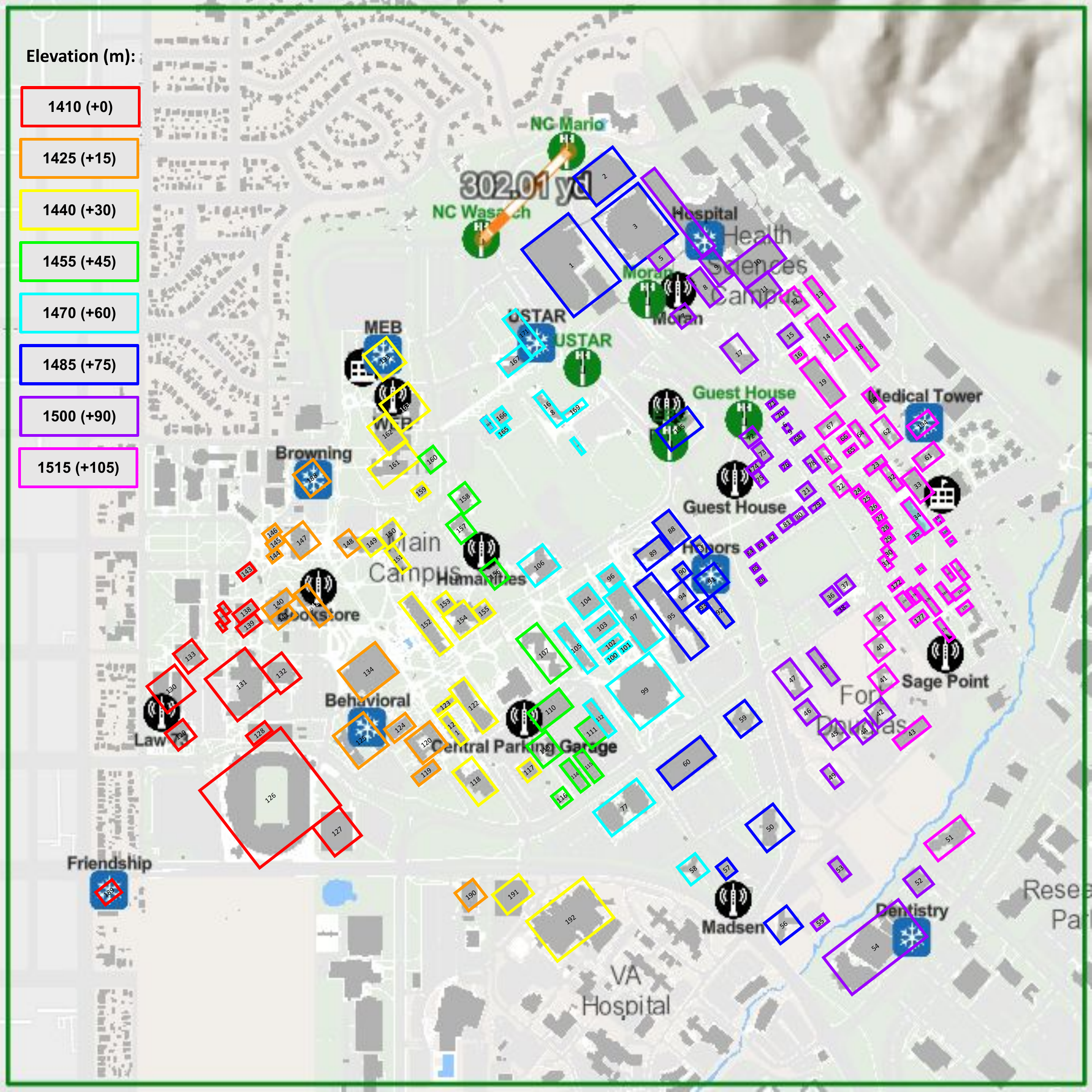}} &
\frame{\includegraphics[width=0.2\textwidth]{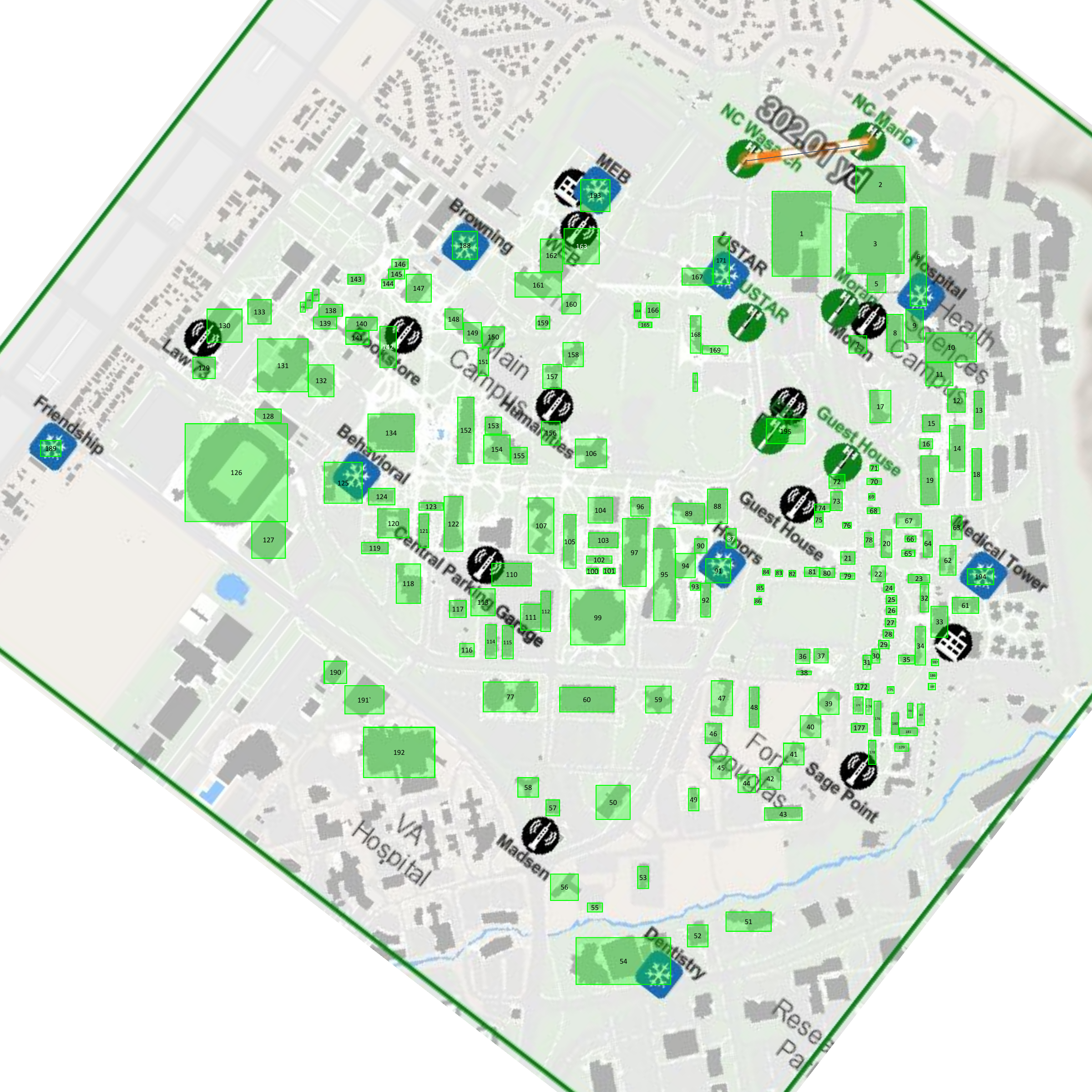}} &
\frame{\includegraphics[width=0.2\textwidth]{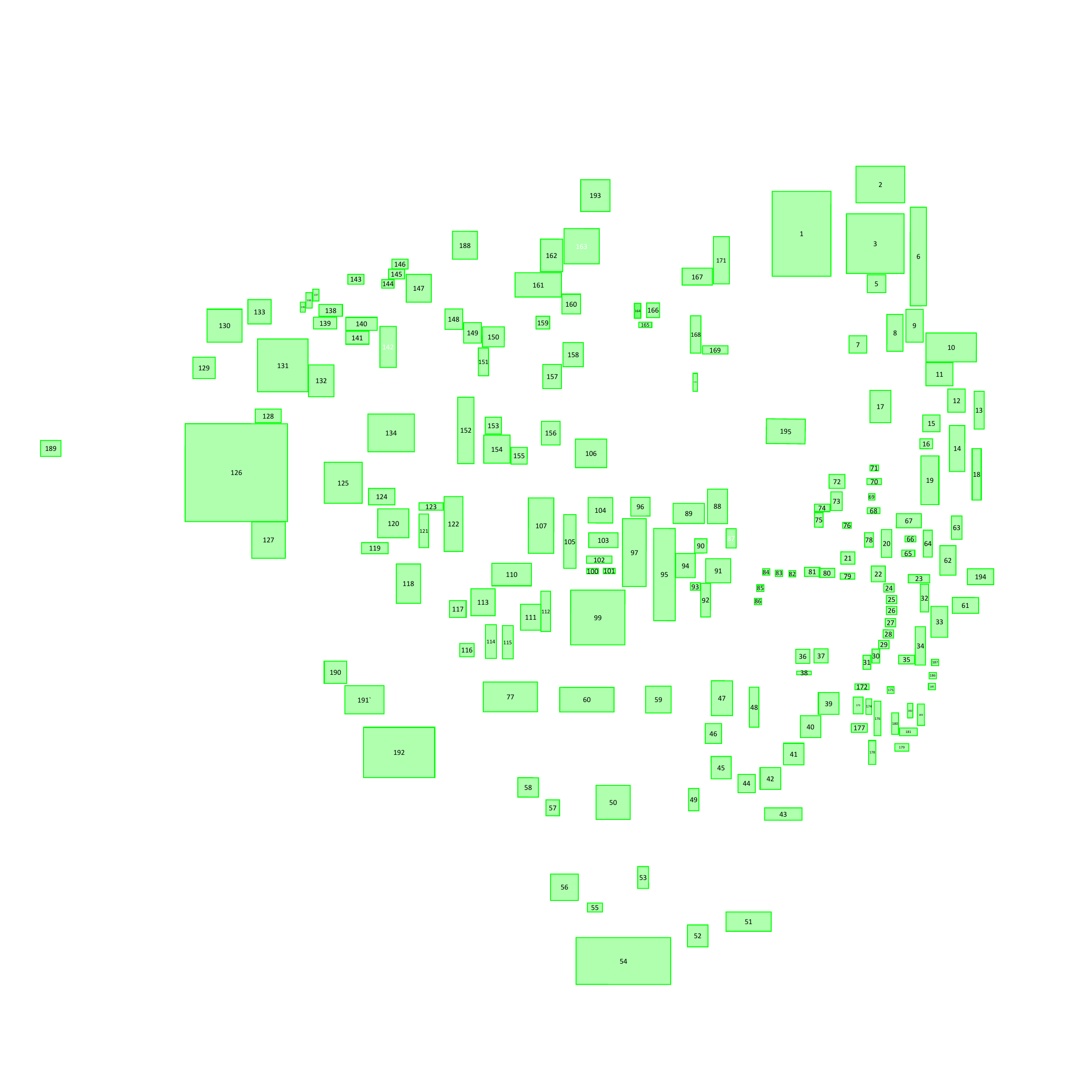}}
% &
% \hspace{-12mm}
\\ (a) & (b) & (c) & (d)  
\end{tabular}
\vspace{-3mm}
\caption{\small Virtualization process of POWDER deployment environment: (a) generate best-fit bounding boxes for all buildings on the University of Utah campus using building dimension data; (b) assign building elevation based on topological data; (c) align bounding boxes with major axis; and (d) isolate AABB objects to import to UBSim.}
% \vspace{-3mm}
\label{fig:virtualization_process}
\end{center}
\end{figure*}

\section{Related Work}\label{sec:related} 
% New applications of DT technology have resulted in increased focus on the DT design process in recent literature. 
% %
% The authors of \cite{testolina2024bostonTwin} create a digital twin of Boston, MA using a high-fidelity 3D model and known cellular base station deployments, and provide a set of APIs for interacting with the 3D model.
Recent literature has seen a growing range of applications of DT technology in wireless systems. For example, the authors of \cite{testolina2024bostonTwin} develop a high-fidelity DT of the city of Boston, incorporating existing cellular base station deployments. 
% They also provide a set of APIs for interacting with this 3D model.
The authors of \cite{villa2024colosseumDT} utilize the Colosseum platform to create a DT of the publicly available OTA indoor testbed Arena \cite{bertizzolo2020arena}. They detail the DT creation process and demonstrate that their framework achieves over 98\% similarity in network performance, while also revealing the sim-to-real gap. In another study, \cite{pinyo2021sim2realtransfer} shows that the sim-to-real gap in a federated edge computing framework is minor, thanks to the high fidelity of trace-based channel modeling. Additionally, \cite{gao2023wiprox} explores sim-to-real transfer learning for device proximity estimation, where a deep learning policy is trained in simulation and fine-tuned in a real network, achieving 93\% prediction accuracy. 

These studies typically focus on testbeds with limited scale or controlled environments, leaving the study of the sim-to-real gap in long-range, outdoor scenarios relatively underexplored. Differently, in this work we examine the effects of model selection on the performance gap between DT simulations and real network performance. 
% We develop a virtualization of the POWDER platform using an open-source wireless network simulator, evaluate the impact of various path loss models on the sim-to-real gap for data-driven modeling and control tasks, and identify critical design considerations to enhance the accuracy and practical implementation of future DT frameworks.
To the best of our knowledge, this work presents the first examination of the sim-to-real gap for data-driven modeling and control within community testbed platforms for advanced wireless research.

\section{Virtual Testbed Design 
% \todo{This paragraph is only one of the design steps}
}\label{sec:design} 

% We have divided the virtual testbed design process into two steps.
% % sections. 
% First, we perform geographical reconstruction of the testbed deployment in the simulation engine,
% % , which has been introduced in Section~\ref{sec:powder}, 
% including 3D geometry of all buildings, local topology, and modeled node types. 
% The second step is behavioral modeling, in which we define parameters and functions of each modeled node type in the simulation engine. 
% % In this step, we consider path loss and signal fading to be key elements of behavior modeling for accurate simulation of wireless communications at the scale presented by the POWDER rooftop node deployments, and implement several well-known path loss models to achieve link-level modeling in the virtual testbed design. 
% % As a key element of this behavior modeling, we formulate a specialized path loss model based on data collected from POWDER  which can predict link performance with greater accuracy than more generalized approaches.
% Before detailing each step, next we first provide a brief overview of the POWDER platform, highlighting the key Components for virtualization.

We have divided the virtual testbed design process into two distinct steps. First, we conduct a geographical reconstruction of the testbed deployment within the simulation engine, incorporating 3D models of buildings, local topology, and node types. The second step involves behavioral modeling, where we define the parameters and functions for each modeled node type in the simulation engine. Before delving into these steps, we provide a brief overview of the POWDER platform, highlighting the key components essential for virtualization.

\subsection{POWDER Platform: A Primer}

POWDER \cite{breen2020powder} is a highly configurable, city-scale software-defined wireless networking testbed managed by the University of Utah in Salt Lake City, Utah. It features dozens of nodes deployed across a 14 km² area around Salt Lake City, including metropolitan, residential, and campus environments. The primary testbed is situated at the University of Utah campus and includes several node types: rooftop nodes, fixed (ground-level) endpoints, densely deployed nodes, mobile endpoints, and mMIMO nodes.  

Given their geographic relevance, user popularity, and availability, we focus on the rooftop node deployments as central elements in the virtual framework design. There are eight rooftop nodes designated for use in the CBRS band, each equipped with two USRP X310 software-defined radios (SDRs) with UBX160 daughterboards and two USRP N310 SDRs with CBRS RF front-ends. All nodes are connected to a remote access and control framework based on Emulab, via various clusters located at the network edge. 
% The overall architecture of the POWDER platform, encompassing resource orchestration and available software, is illustrated in Fig.~\ref{fig:powder_architecture}.
% This setup facilitates direct hardware access through edge servers, minimizing latency and maximizing configurability during remote operations, as well as central resource scheduling via the POWDER website.

\subsection{Virtual Environment Construction}

% In order to construct a virtual replica of the POWDER deployment area, geographical information system (GIS) data was provided from both the University of Utah Department of Geography as well as collected from the U.S. Geological Survey National Map \cite{utah_gis_map}. Among the data collected include building height and elevation measurements, aerial photography of the University of Utah campus, and GPS coordinates of selected POWDER nodes. 

To construct a virtual replica of the POWDER deployment area, we utilized geographical information system (GIS) data obtained from the University of Utah Department of Geography and the U.S. Geological Survey National Map \cite{utah_gis_map}. The collected data includes building height and elevation measurements, aerial photography of the University of Utah campus, and GPS coordinates for selected POWDER nodes.

% The virtual system construction process is shown in Fig.~\ref{fig:virtualization_process}, where the geographic information of all nodes, 
% % introduced in Section~\ref{sec:powder}, 
% including GPS coordinates and height data, was provided directly by the POWDER Team at the University of Utah. These coordinates correspond directly to the nodes observable via the POWDER Live Map \cite{powder_live_map}.
% % as shown in Figure~\ref{fig:livemap}. 
% The rooftop nodes are represented on this map as blue snowflake icons, and the ground-level endpoints are represented as black antenna icons. 
% As mentioned above,
% % in Section~\ref{sec:powder}, 
% these node types are of particular interest to the sim-to-real gap measurement process. 
% The Dense Deployment nodes are indicated by the green antenna icons, and the Data Centers indicated by the black building icon provide edge compute functionality and do not provide radio hardware as a part of the testbed.

The process of constructing the virtual system is illustrated in Fig.~\ref{fig:virtualization_process}. 
% The geographic information, including GPS coordinates and height data, was provided by the POWDER Team at the University of Utah. These coordinates correspond directly to the nodes visible on the POWDER Live Map. 
On these maps, rooftop nodes are depicted as blue snowflake icons, while ground-level endpoints are shown as black antenna icons. 
% These node types are particularly relevant for measuring the sim-to-real gap. 
Dense deployment nodes are marked with green antenna icons, and data centers are represented by black building icons, offering edge computing functionality.
% but do not include radio hardware as part of the testbed.

% We define the properties and behavior of each selected node type in the NEM of the simulation engine, modeling physical properties such as number of antennas and hardware dimensions as well as behavioral properties such as transmit power, modulation, frequency and bandwidth.  
% In order to import the POWDER geographical information to UBSim, we first begin by generating best-fit axis-aligned bounding boxes (AABB) using a map of buildings within the range of the on-campus POWDER deployment. The bounding boxes provide a general layout of the campus to establish line-of-sight (LOS) or non-line-of-sight (NLOS) between deployed nodes.
% As shown in Figure~\ref{fig:virtualization_process}(a), the majority of buildings on the University of Utah campus are geometrically aligned and can be directly represented as AABB objects, while some non-axis-aligned buildings require geometric compensation. 
% Building height data was provided directly by the University of Utah. 
% The next step was to import elevation data for each building from the USGS National Map \cite{utah_gis_map}. 
% The elevation map was quantized into steps of 15 m based on building groupings, which are represented as different colors in Figure~\ref{fig:virtualization_process}(b). 
% The reference map was then rotated to align with the major axis used in UBSim as shown in Figure~\ref{fig:virtualization_process}(c), and finally Figure~\ref{fig:virtualization_process}(d) shows the isolated AABB map objects which were imported to the Environment Definition API. 

We use the simulation engine UBSim \cite{NEXTCOMCOM} to model the behaviors of each selected node type. To import the POWDER geographical information into UBSim, we first generate best-fit axis-aligned bounding boxes (AABBs) using a map of buildings within the on-campus POWDER deployment area. These bounding boxes provide a general layout of the campus, allowing us to determine line-of-sight (LOS) or non-line-of-sight (NLOS) conditions between deployed nodes. As illustrated in Fig.~\ref{fig:virtualization_process}(a), most buildings on the University of Utah campus are geometrically aligned and can be represented directly as AABBs, while some non-axis-aligned buildings require geometric adjustments. Building height data was supplied by the University of Utah. Next, we import elevation data for each building from the USGS National Map \cite{utah_gis_map}. The elevation data is quantized into 15-meter intervals based on building groupings, as depicted by different colors in Fig.~\ref{fig:virtualization_process}(b). The reference map is then rotated to align with the major axis used in UBSim, as shown in Fig.~\ref{fig:virtualization_process}(c). Finally, Fig.~\ref{fig:virtualization_process}(d) displays the isolated AABB map objects, as imported into the Environment Definition API.
In this work, we characterize the behaviors of select nodes based on both OTA measurements and standardized path loss models, as discussed below. 

\subsection{OTA Measurement}
% The OTA measurement is conducted using the Shout measurement framework \cite{Webb2021Shout}, which is an open-source point-to-point measurement framework provided by default in several POWDER profiles to collect SINR for both the rooftop and ground-level scenarios. 
% Shout performs orchestrated wireless measurements via signal broadcast to collect link state information including RSSI and SINR during signal transmission, and can automate the TX/RX functions across multiple POWDER nodes. 
% This framework provides two default waveform options for over-the-air measurement: constant-waveform (CW) transmission, and direct-sequence spread spectrum (DSSS) modulation. 
% Due to the interference levels present during several rounds of measurement, we elect to use the DSSS transmission method. 

The OTA measurements are performed using the Shout measurement framework \cite{Webb2021Shout}, an open-source point-to-point measurement tool included by default in various POWDER profiles. 
% Shout facilitates orchestrated wireless measurements by broadcasting signals to collect link state information, including RSSI and SINR, during signal transmission. It also automates the transmission and reception functions across multiple POWDER nodes. 
The framework offers two default waveform options for OTA measurement: constant-waveform (CW) transmission and direct-sequence spread spectrum (DSSS) modulation. Given the interference levels observed during several measurement rounds, we have chosen to use the DSSS transmission method in our OTA measurements.
DSSS is a spread-spectrum modulation technique which is used to provide both good signal quality in the presence of interference as well as data security by spreading transmitted information over a large bandwidth. To achieve this spreading, each modulated symbol of transmitted signal $x$ with transmit power $P_{t}$ is multiplied with a pseudo-random noise (PN) sequence, termed the spreading code $\mathbf{c}$, resulting in a sequence of chips. 

As detailed in Section~\ref{sec:design}, our investigation focuses on communication between rooftop node deployments. To design these scenarios, we considered a comprehensive set of point-to-point link combinations among all available rooftop nodes, ensuring a robust experimental campaign. Data collection for each combination was carried out using the \textit{``shout-long-2024"} profile, a customized configuration built around the Shout framework. This profile specifies transmission parameters, including PN sequence length, transmitter and receiver gain, measurement duration and repetition, and node selection, all organized into JSON files for straightforward user configuration. 
% The JSON configuration is disseminated among all connected nodes via an orchestrator node.

% Similar profiles and instructions for use and customization are available to the community through the POWDER platform. 
Measurements were conducted opportunistically in the newly opened CBRS band, as outlined in Section~\ref{sec:intro}, by first sensing available spectrum resources to avoid interference. All measurements were performed within a center frequency range of 3550 to 3700 MHz, with minimal variation in frequency based on resource availability. It is worth noting that we have observed negligible differences in link performance within this frequency range. 
% Additionally, we ensured optimal signal strength by transmitting and receiving at maximum gain across all links in each scenario.

% \todo{Try to link to previous sections}
% We consider a round-robin transmission scheme to measure link performance between all rooftop-to-rooftop links. 
% The scenario is prepared by reserving a number of rooftop nodes, introduced in Section~\ref{sec:design}, and instantiating the selected profile at the time of reservation. One compute node is reserved in addition, which serves as the orchestrator node to coordinate automatic send-and-receive operations among the selected transceivers. 
% To collect measurements for this scenario, the selected rooftop node will transmit a DSSS signal to all other selected rooftop nodes. 
% This transmission will be repeated 10 times, and measurements of average RSSI and SINR for this set of transmissions will be stored at each receiving rooftop node. 
% These measurements will be collected by the orchestrator, and compiled into an HDF5 file for processing. 
% Following this measurement step, the next rooftop node will be selected as the transmitter and all other rooftop nodes will be designated as receivers, and the same procedure is executed until all possible links have been evaluated. The rooftop nodes evaluated for this experiment were the Friendship, Browning, Behavioral (BES), Honors, USTAR, Hospital, and Merrill Engineering Building (MEB) nodes.
% the locations of which are indicated in Figure~\ref{fig:livemap}. 
% The Dentistry node was unavailable at the time of this experiment.  

We employ a round-robin transmission scheme to assess link performance between all rooftop-to-rooftop links. The scenario is set up by reserving a subset of rooftop nodes, as introduced in Section~\ref{sec:design}, and configuring the selected profile at the time of reservation. Additionally, one compute node is reserved to act as the orchestrator, coordinating the automatic transmission and reception operations among the selected transceivers.
To gather measurements, each selected rooftop node transmits a DSSS signal to all other selected rooftop nodes. This transmission is repeated 10 times, with average RSSI and SINR measurements collected at each receiving rooftop node. These measurements are then aggregated by the orchestrator and compiled into an HDF5 file for further processing.

Following this, the next rooftop node is designated as the transmitter, and the procedure is repeated with all other rooftop nodes serving as receivers. This process continues until all possible links have been evaluated. The rooftop nodes involved in this experiment include the Friendship Manor (FM), Browning, Behavioral (BES), Honors, USTAR, Hospital, and Merrill Engineering Building (MEB) nodes. The Dentistry node was unavailable at the time of this experiment.

% For a simple DSSS signal s(t), 
% \begin{align}
%     s(t) = A_m(t) * c(t) *cos(2bf_ct)
% \end{align}
% where $A_m(t)$ is the modulated signal, which is the original signal, $c(t)$ is the spreading code with a chip rate $R_c$, $b$ is the gold cold length, and $f_c$ is the carrier frequency.

% % Autocorrelation
% \begin{align}
%     R(k) = (a * b)(k-N+1) = \sum_{l = 0}^{\|a\|-1} a_l b_{l-k+N-1} 
% \end{align}
% where $a$ is the phase rotation and $N$ is the length of the spreading code in chips.

% The spreading gain of a DSSS signal is
% \begin{align}
%     \text{Spreading gain} = 10 log_{10} * \text{spreading factor}
% \end{align}
% where spreading factor = $\frac{R_c}{N}$. This spreading factor is used to determine the bandwidth $B = N * \text{spreading factor}$ of a DSSS signal.

\subsection{Standardized Path Loss Models}

Several standardized path loss models have been investigated to deploy in the simulated POWDER environment to accurately portray the RF conditions at the University of Utah propagation environment. These include the Friis free space path loss model, the ECC-33 model, and the Ericsson model.
% \cite{mollel2014pathloss}. 
% These models were selected due to their generalization capability and prediction accuracy in urban environments. 
These models were chosen for their proven generalization capabilities and prediction accuracy in urban settings.

% The free space model is widely used to estimate path loss for a given frequency as a function of distance, considering different forms for LOS and NLOS links, and was selected 
% % due to 
% because of
% its prevalence and ubiquity in RF scenarios. The model is formulated as follows, where $d$ is the link distance in km and $f$ is the transmission frequency in MHz: 
The free-space path loss model is commonly employed to estimate path loss based on frequency and distance, accounting for both line-of-sight (LOS) and non-line-of-sight (NLOS) conditions. Its widespread use in RF scenarios makes it a suitable choice for our analysis. The model is defined as $PL(dB) = 32.45 + 20 \log_{10}(d) + 20 \log_{10}(f)$, where $d$ represents the link distance in kilometers and $f$ denotes the transmission frequency in megahertz.
% \begin{equation}
%     PL(dB) = 32.45 + 20 \log_{10}(d) + 20 \log_{10}(f). 
% \end{equation}

% The ECC-33 model is an extension of the Okumura and Hata path loss models to include frequencies up to 8 GHz. This model improves on the free space model by incorporating gains based on base station height and receiver height, and is formulated in \cite{abhaya2005models} as follows: 
The ECC-33 model extends the Okumura and Hata path loss models to accommodate frequencies up to 8 GHz. It enhances the free-space model by incorporating gains related to base station and receiver heights. The model is formulated as $PL(dB) = A_{fs} + A_{bm} - G_{b} - G_{r}$, where 
% \begin{equation}
%     PL(dB) = A_{fs} + A_{bm} - G_{b} - G_{r}
% \end{equation}
% in which 
$A_{fs}$ is the free space attenuation, $A_{bm}$ is the basic median path loss, $G_{b}$ is the base station (BS) height gain factor, and $G_{r}$ is the receiver height gain factor. 
% These values are defined for ``medium city" environments as: 
% \begin{equation}
%     A_{fs} = 92.4 + 20 \log_{10}(d) + 20 \log_{10}(f)
% \end{equation}
% \begin{equation}
% \begin{split}
%     A_{bm} = & 20.41 + 9.83 \log_{10}(d) + 7.894 \log_{10}(f) + \\
%              & 9.56 [\log_{10}(f)^2]
% \end{split}
% \end{equation}
% \begin{equation}
%     G_{b} = \log_{10}(\frac{h_{b}}{200}) * [13.958 + 5.8[\log_{10}(d)^2]]
% \end{equation}
% \begin{equation}
%     G_{r} = [42.57 + 13.7 \log_{10}(f)] * [\log_{10}(H_{r}) - 0.585]
% \end{equation}
% \begin{align}
%     A_{fs} = \; & 92.4 + 20 \log_{10}(d) + 20 \log_{10}(f)\\
%     A_{bm} = \; & 20.41 + 9.83 \log_{10}(d) + 7.894 \log_{10}(f) \nonumber \\
%  & +  9.56 [\log_{10}(f)^2]\\
%    G_{b} = \; & \log_{10}(\frac{h_{b}}{200}) * [13.958 + 5.8[\log_{10}(d)]^2]\\
%    G_{r} =\;  & [42.57 + 13.7 \log_{10}(f)] * [\log_{10}(h_{r}) - 0.585]
% \end{align}
% in which $h_b$ is the height of the base station in meters and $h_r$ is the height of the receiver in meters. 

% Finally, the Ericsson model is an extension of the ECC-33 model which can provide further consideration of deployment environment. This model is formulated in \cite{mollel2014pathloss} as follows: 
Finally, the Ericsson model builds upon the ECC-33 model by providing additional considerations for deployment environments. This extension allows for a more detailed representation of various environmental factors. The model is formulated as follows:
\begin{equation}
\begin{split}
        PL(dB) = & a_{0} + a_{1}\log_{10}(d) + a_{2}\log_{10}(h_{b}) + \\
                 & a_{3}\log_{10}(h_{b})\log_{10}(d) - \\
                 & 3.2 (\log_{10}(11.75 h_{r})^2)+g(f) 
\end{split}
\end{equation}
where the frequency-dependent component $g(f)$ is defined as $g(f) = 44.49\log_{10}(f) - 4.78(\log_{10}(f)^2)$, 
% \begin{equation}
%     g(f) = 44.49\log_{10}(f) - 4.78(\log_{10}(f)^2),
% \end{equation}
and the coefficients $a_{0}, a_{1}, a_{2},$ and $a_{3}$ are set for urban environments with values of $36.2, 30.2, 12,$ and $0.1$, respectively\footnote{It is found in our experiments that these typical coefficients for urban, suburban, and rural environments do not fit the collected data due to the topographical variation of the physical environments. Therefore, we adopted $a_{0} = 26, a_{1} = 10, a_{2} = 12,$ and $a_{3} = 0.1$ to enhance model accuracy based on the collected data.}.

\subsection{Synthetic Dataset Generation}

To generate the synthetic training data, we implement the propagation models detailed in Sec.~\ref{sec:design} in the virtual POWDER environment deployed in UBSim to take advantage of its 3D geometry. The simulator was configured to replicate the experiment described earlier in this section, estimating SINR and BER for each transmitter-receiver pair. The goal of this process is to match the OTA data collected from the POWDER testbed, using link distance, transmitter and receiver height, spread spectrum processing gain, and LOS conditions from the virtual model. 
This results in three synthetic datasets, one for each propagation model considered.
% included in the scope of this work. 
%
% As mentioned in Section~\ref{sec:design}, the ``Dentistry" rooftop node was excluded from data collection due to hardware maintenance during the experiment. 

% \setcounter{figure}{2}
\begin{figure}[t]
\begin{center}
\includegraphics[width=0.4\textwidth]{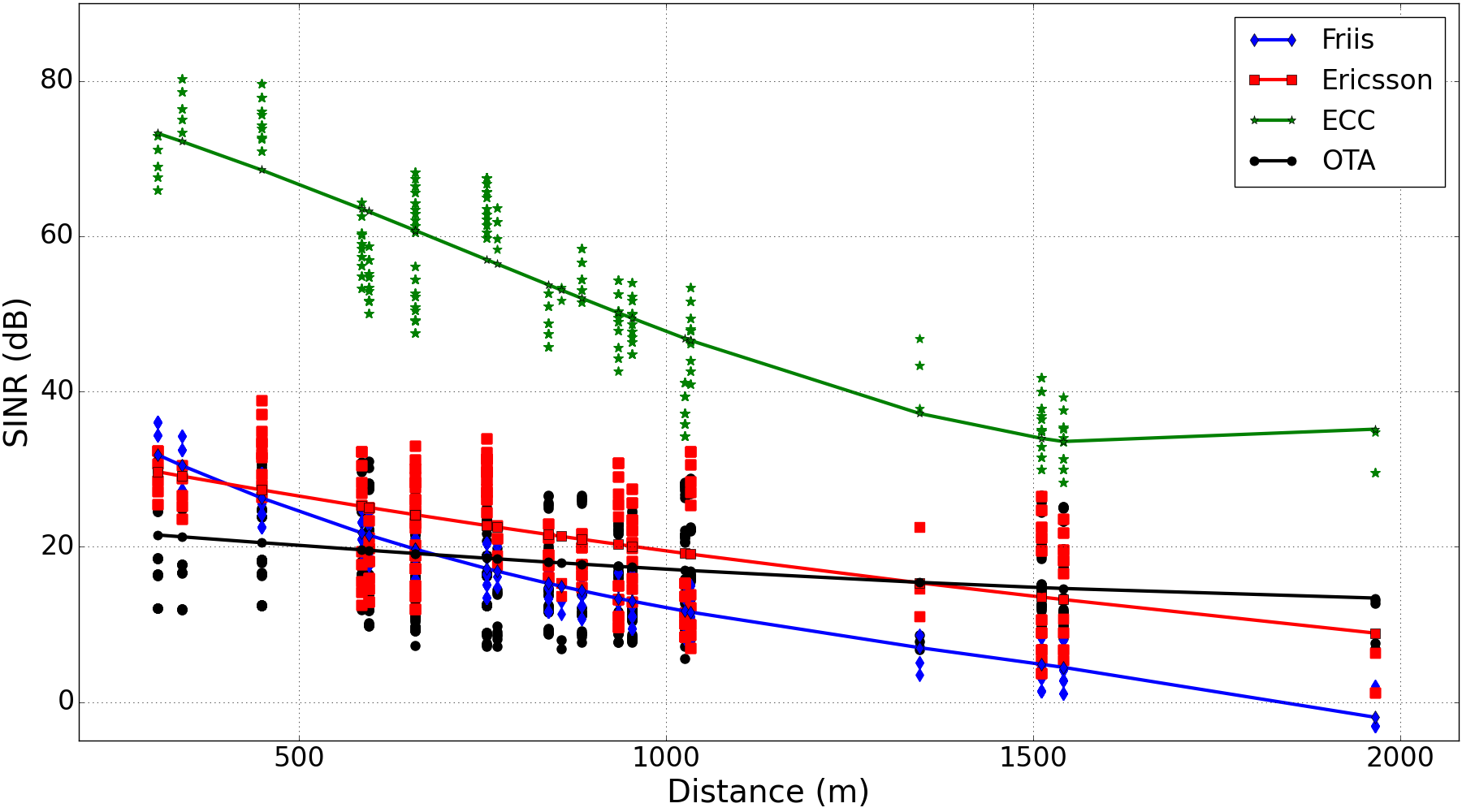}
\caption{\small 
% Comparison of SINR values generated by selected path loss models with values collected from OTA data (black). 
Comparison of SINR values predicted by selected path loss models against values collected from OTA data (black).
}
\vspace{-5mm}
\label{fig:sinrdist}
\end{center}
\end{figure}

% A comparison of the SINR values from each synthetic dataset with the SINR values collected from the OTA data is shown in Fig.~\ref{fig:sinrdist}. A line of best fit has been included for each model. It is clear from this figure that the Ericsson model (red) is capable of generating the most similar estimation of SINR compared to the OTA data. The Friis model (blue) slightly overestimates path loss at greater distance resulting in SINR estimation slightly lower than OTA values, and the ECC model underestimates path loss, resulting in excessively optimistic SINR values. While the ECC model is known to typically overestimate path loss, 
% % as in \cite{mollel2014pathloss} and \cite{mollel2014pathloss}, 
% the gain terms associated with transmitter and receiver height discussed in Sec.~\ref{sec:design} may not account for the steep elevation changes of the University of Utah campus which are added to the node height measurements during the 3D modeling process. 

A comparison of SINR values from each synthetic dataset against the OTA data is presented in Fig.~\ref{fig:sinrdist}, with a line of best fit included for each model. It can be seen that the Ericsson model (red) produces the most accurate SINR estimates, closely aligning with the OTA data. In contrast, the Friis model (blue) slightly overestimates path loss at greater distances, leading to lower-than-expected SINR values. The ECC model, however, underestimates path loss, resulting in overly optimistic SINR predictions. Although the ECC model tends to overestimate path loss, the gain terms related to transmitter and receiver height may fail to fully account for the steep elevation changes on the University of Utah campus, which are incorporated during the 3D modeling process.

\section{Data-driven Policy Selection}\label{sec:datadriven}

% As outlined in Sec.~\ref{sec:intro}, we outline the data-driven models which we will use to demonstrate the sim-to-real gap induced by the models outlined in Sec.~\ref{sec:design}. For this task, we design two applications which are representative of general trends of how data-driven models are applied the surveyed literature: predictive modeling of network behaviors, and network protocol optimization. 
% We note that for the remainder of the work, bold-faced variables refer to vectors or matrices
As discussed in Section~\ref{sec:intro}, this work aims to investigate the impact of the sim-to-real gap on data-driven control in wireless networks. 
% introduce the data-driven models used to demonstrate the sim-to-real gap caused by the models described in Section~\ref{sec:design}. F
To this end, we consider two representative networking tasks: predictive modeling of network behavior and network protocol optimization. 
% These applications serve to highlight the impact of the sim-to-real gap on typical data-driven tasks in wireless networks.

\begin{figure}[t]
\begin{center}
\includegraphics[width=0.35\textwidth]{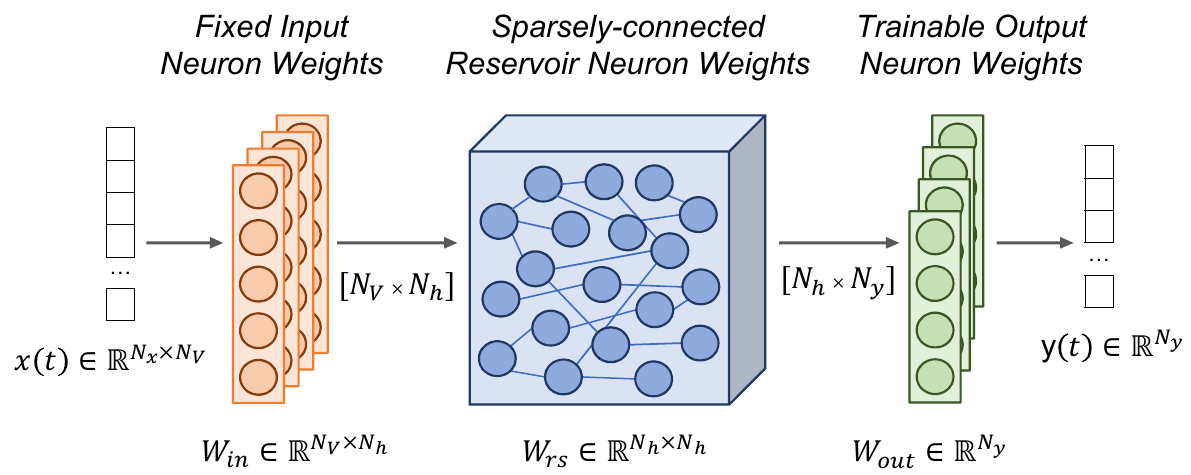}
\caption{\small ESN Architecture. 
}
\vspace{-6mm}
\label{fig:esn_arch}
\end{center}
\end{figure}

For both tasks, we leverage an Echo State Network (ESN) \cite{bianchi2021esn} for policy training. 
% ESN is a type of recurrent neural network based on reservoir computing, to serve as the data-driven model and control policy, respectively. 
ESN is a type of recurrent neural network based on reservoir computing. As shown in Fig.~\ref{fig:esn_arch}
the ESN architecture is comprised of an input layer, a reservoir, and a trainable output layer.
% To generalize the discussion, we 
Define the ESN learning process with the number of observed variables $N_{V}$, the number of observed time steps, $N_{t}$, the number of collected data points, $N_{x}$, the size of the ESN reservoir $N_{h} \times N_{h}$, and the number of output variables, $N_{y}$. 
Then, the reservoir can be implemented as a series of neurons which define the internal state, represented as vector of activation outputs, $\mathbf{h}(t)\in\mathbb{R}^{N_{h}}$, to be a function of the ESN input. In most cases, the neuron model provides some leakage parameter, $\alpha$, by which the activation is determined by both $\mathbf{h}(t)$ as well as previous state $\mathbf{h}(t-1)$, which improves learning capability of temporal features from the input data. 
However, neither the input layer weights $\mathbf{W}_{in}\in\mathbb{R}^{N_{V} \times N_{h}}$ nor the reservoir weights $\mathbf{W}_{rs}\in\mathbb{R}^{N_{h} \times N_{h}}$ are trainable. 
The only trainable weights in the ESN are those of the output layer, $\mathbf{W}_{out}\in\mathbb{R}^{N_{y} \times (1 + N_{V} + N_{h})}$.

\begin{figure*}[t]
\begin{center}
\begin{tabular}{cc}
\includegraphics[width=0.38\textwidth]{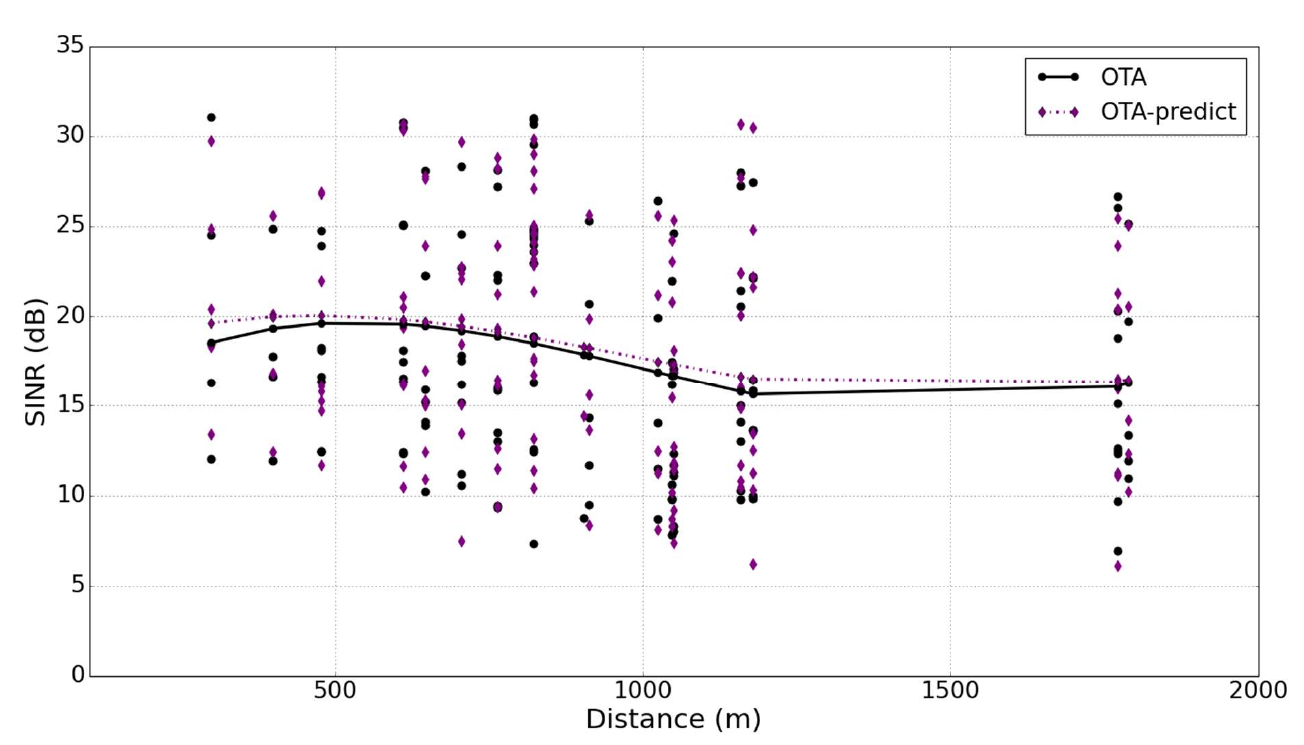} &
\includegraphics[width=0.38\textwidth]{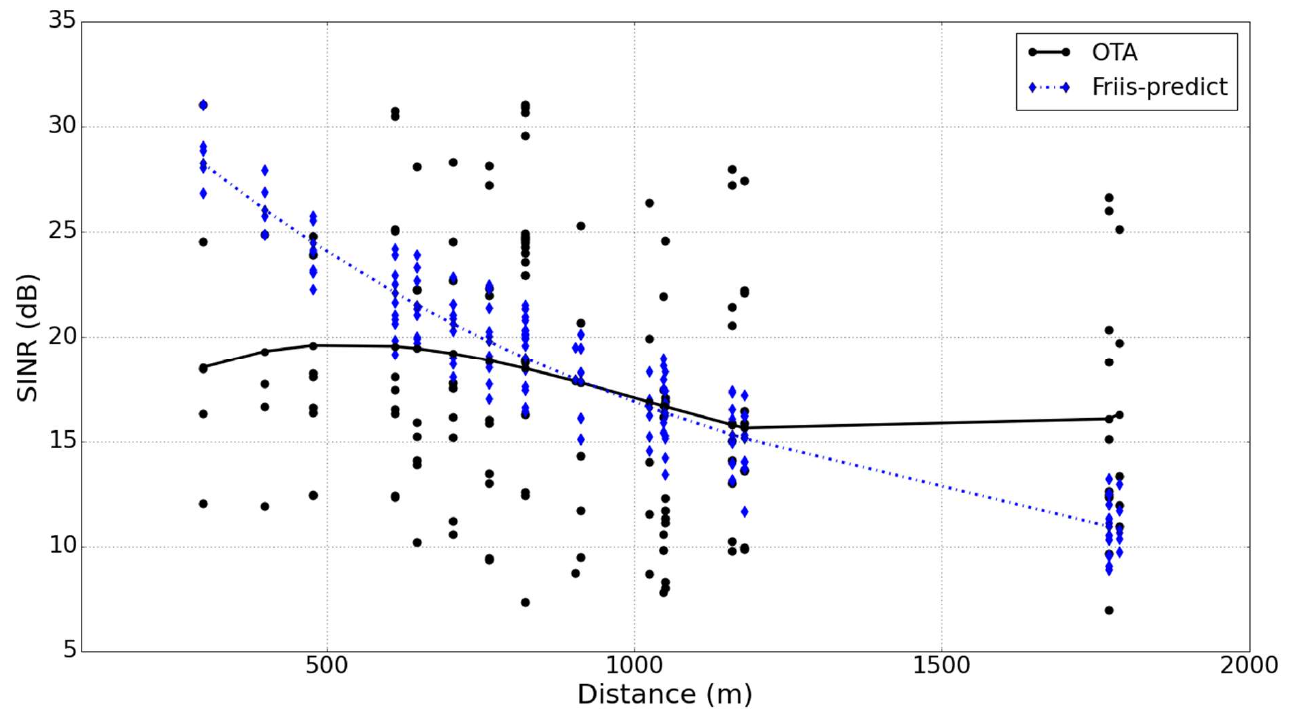}  \\
(a) & (b) \\ 
\includegraphics[width=0.38\textwidth]{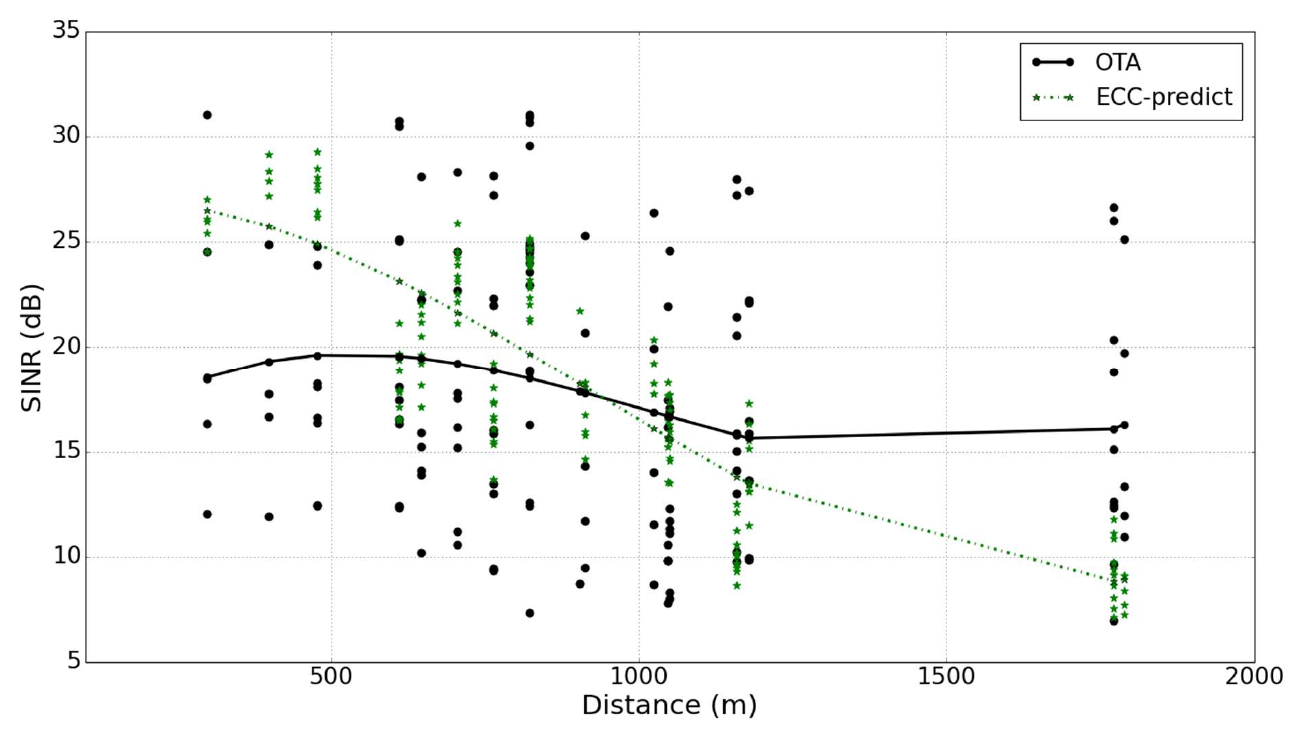} &
\includegraphics[width=0.38\textwidth]{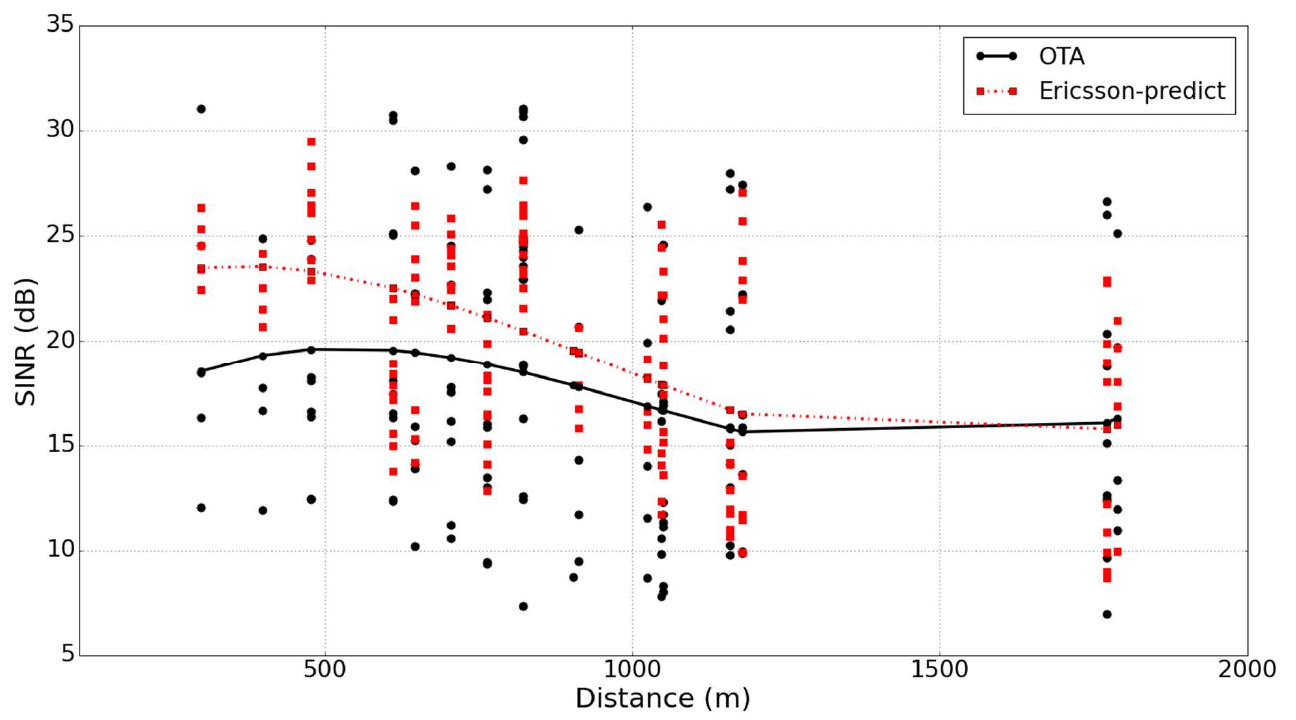} \\
% &
% \hspace{-12mm}
(c) & (d)  \\
\end{tabular}
\caption{\small Prediction of point-to-point SINR between rooftop nodes for the trained models: (a) ESN-1-OTA model (real-to-real), (b) ESN-1-Friis model (sim-to-real), (c) ESN-1-ECC model (sim-to-real), and (d) ESN-1-Ericsson model (sim-to-real).}
\vspace{-5mm}
\label{fig:sinr_pred_plots}
\end{center}
\end{figure*}

In order to apply the ESN to our tasks without significant temporal dependencies among inputs, a known output $\mathbf{y}^{*}(t-1)\in\mathbf{R}^{N_{y}}$ can be applied in place of $\mathbf{y}(t-1)$ to remove the recurrent relationship between $\mathbf{h}(t)$ and $\mathbf{y}(t)$. This approach is termed \textit{teacher forcing}, and can both accelerate the training process and adapt the network to more generalized learning tasks. For both tasks, we leverage teacher forcing to enable single-time-step evaluation since neither task relies on temporal features from the collected data. We use the PyESN library \cite{pyESNgithub}, and 
% follow recommendations outlined in \cite{Lukosevicius2012esn} 
leverage a standard grid search method
for hyperparameter selection.  

% \subsection{Task 1: Predictive Modeling}

% The selected predictive modeling task is the estimation of 
\textit{Task 1: Predictive Modeling:} The objective of this task is to estimate the signal-to-interference-plus-noise ratio (SINR) for a given link and protocol configuration. This task has been selected due to its ubiquity in existing literature, and is consistently leveraged as a basis for 
% network deployment optimization \cite{borralho2024celldeployment}, link capacity estimation in various spectrum bands \cite{van2022iiot}, and optimizing spectrum coexistence and device co-location \cite{bonati2021colosseum}. 
%
network deployment optimization, link capacity estimation in various spectrum bands, and optimizing spectrum coexistence and device co-location. 
For this task, we set $N_{V}=8$, using variables such as TX-RX distance, TX node height, RX node height, and Gold code length, which has been one-hot encoded into five separate variables corresponding to the available options: 31-bit, 63-bit, 127-bit, 511-bit, and 2047-bit Gold code lengths. We set $N_{y}=1$, representing the estimated SINR.

% \subsection{Task 2: Protocol Optimization}

\textit{Task 2: Protocol Optimization:} 
% The protocol optimization task is the estimation of 
This task aims to predict the optimal PN code length for a given network topology and observed conditions. Due to the potential concerns for spectrum congestion while using the POWDER platform outlined in Sec.~\ref{sec:intro}, this task is intended to demonstrate how the use of a DT can improve communications performance through protocol configuration. Take advantage of the extensive parameterization of the Shout framework, the PN code can be selected to effectively balance the inherent tradeoff between resilience and throughput associated with DSSS PN code length selection  outlined in Sec.~\ref{sec:design}.
For this task, we set $N_{V}=5$, specifically using TX-RX distance, TX node height, RX node height, observed SINR, and observed BER.  
%
% We set $N_{y}=1$, corresponding to each of the available Gold code lengths, which have been one-hot encoded as mentioned above. 
We set $N_{y}=5$, corresponding to each of the available one-hot encoded Gold code lengths.
% , which have been one-hot encoded as mentioned above.

% \section{Experimental Framework and Results}\label{sec:framework} 
\section{Experimental Evaluation and Results}\label{sec:evaluation}

% In order to demonstrate the effects of the sim-to-real gap induced by the virtual POWDER testbed environment, we train a set of data-driven policies to accomplish the tasks outlined in Sec.~\ref{sec:datadriven}. For each task, we split the OTA dataset into training and evaluation datasets, then train a ground truth model on the OTA training data. Then, for each synthetic dataset, we train two models, one for each task. Finally, all trained models are evaluated on the reserved OTA evaluation dataset. The performance of each model trained on the synthetic datasets is compared with the OTA model for the associated task to demonstrate the sim-to-real gap induced by each considered propagation model. 

To illustrate the impact of the sim-to-real gap caused by the virtual POWDER testbed environment, we train a series of data-driven policies to address the tasks described in Sec.~\ref{sec:datadriven}. We split the OTA dataset into training and evaluation sets, training a ground truth model for each task on the OTA training data. For each synthetic dataset, we train two models—one for each task. All models are then evaluated using the reserved OTA evaluation dataset. Finally, we compare the performance of the models trained on synthetic data with the OTA model for each task, highlighting the sim-to-real gap introduced by each propagation model.

% Towards this goal, we design and implement four ESN models to predict the optimal gold code for each collected dataset. The naming convention associates each ESN with the task for which it has been trained as well as the data generation method used to create the training dataset, i.e. \textit{ESN-\#-\{model\}}. We compare the performance of each model by plotting the resulting model outputs alongside the ground truth data, applying a line of best fit to emphasize the relationship between the compared outputs, and calculate Mean Absolute Error (MAE) for each model output. 

To achieve this goal, we design and implement four ESN models to predict the optimal Gold code for each dataset. Each ESN model is named based on the task it was trained for and the data generation method used, following the convention \textit{``ESN-\#-{model}"}. We evaluate the performance of each model by comparing its outputs to the ground truth data, visualizing the results with best-fit lines to highlight the correlation between the outputs, and calculating the Mean Absolute Error (MAE) for each model.

\begin{figure*}[t]
\begin{center}
\begin{tabular}{cccc}
\includegraphics[width=0.22\textwidth]{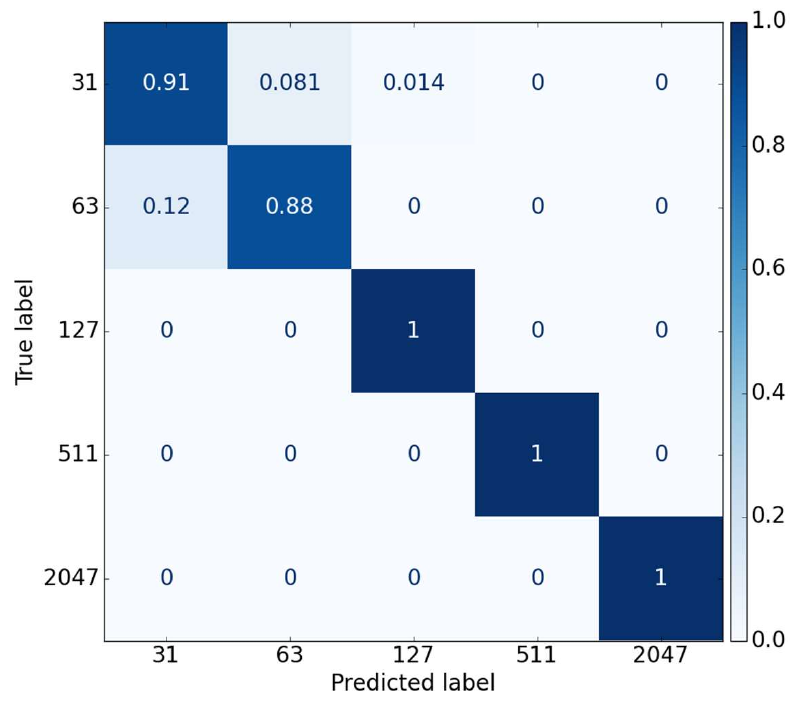} &
\includegraphics[width=0.22\textwidth]{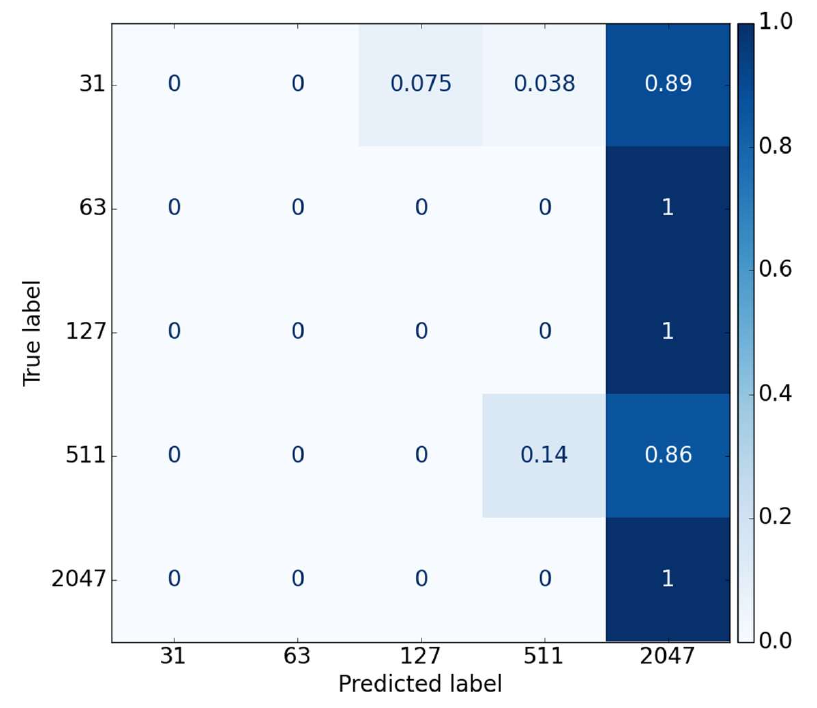} &
\includegraphics[width=0.22\textwidth]{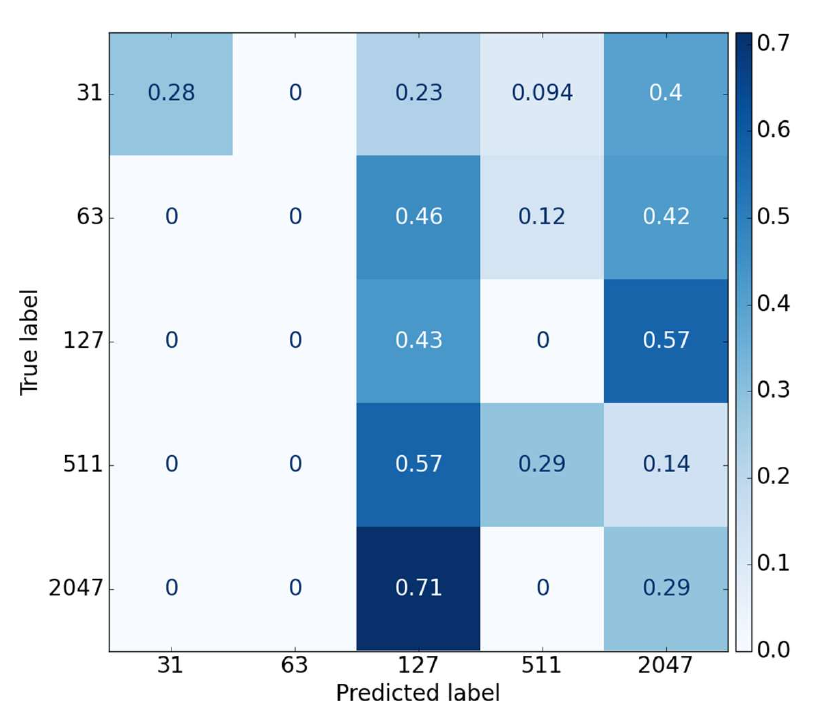} &
\includegraphics[width=0.22\textwidth]{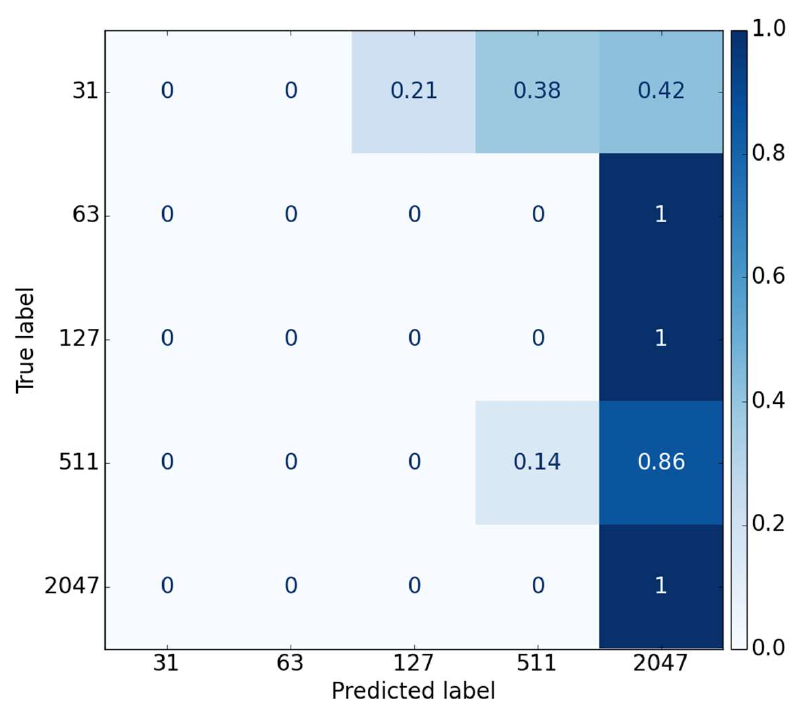}
% &
% \hspace{-12mm}
\\ (a) & (b) & (c) & (d)  
\end{tabular}
\vspace{-2mm}
\caption{\small Prediction of the optimal Gold code length on generated datasets: (a) ESN-2-OTA model (real-to-real), (b) ESN-2-Friis model (sim-to-real), (c) ESN-2-ECC model (sim-to-real), and (d) ESN-2-Ericsson model (sim-to-real).}
\vspace{-6mm}
\label{fig:gold_pred_accuracy}
\end{center}
\end{figure*}

The performance of models trained to predict point-to-point SINR (i.e. Task 1) are shown in Fig.~\ref{fig:sinr_pred_plots}. For each plot, the ground truth SINR results collected from the POWDER testbed are shown in black using a solid line, and each of the compared models are shown in color using a dashed line. 
The results clearly demonstrate the sim-to-real gap on the selected models. 
The ground truth model (i.e. trained on OTA data) is shown in Fig.~\ref{fig:sinr_pred_plots}(a). While the predicted data points show some variance between predicted and ground truth values, the model achieves a MAE of 2.011, and the trend of the prediction data is very similar to the ground truth dataset.  
The Friis model is shown in Fig.~\ref{fig:sinr_pred_plots}(c), and achieves a MAE of 4.957. This model does not incorporate any knowledge of transmitter or receiver height and thus poorly captures the trend of the propagation environment. 
The greatest sim-to-real gap lies with the ECC model, shown in Fig.~\ref{fig:sinr_pred_plots}(c), which achieves a MAE of 5.643.
The Ericsson model, shown in Fig.~\ref{fig:sinr_pred_plots}(c), offers poor performance at lower distances, but provides a good estimation of the trend of ground truth data at larger distances ($\geq$1 km), and achieves a MAE of 5.047. 
We can see that the predictive models based on ECC and Friis path loss models both seem to exaggerate the importance of distance on the resulting path loss in this environment.

% \begin{figure}[t]
% \begin{center}
% \includegraphics[width=0.45\textwidth]{figs/sinr_ota.pdf}
% \caption{\small Comparison of SINR values generated by selected path loss models with values collected from OTA data (black). 
% }
% % \vspace{-4mm}
% \label{fig:results_sinr_ota}
% \end{center}
% \end{figure}

% \begin{figure}[t]
% \begin{center}
% \includegraphics[width=0.45\textwidth]{figs/sinr_friis.pdf}
% \caption{\small Comparison of SINR values generated by selected path loss models with values collected from OTA data (black). 
% }
% % \vspace{-4mm}
% \label{fig:results_sinr_friis}
% \end{center}
% \end{figure}

% \begin{figure}[t]
% \begin{center}
% \includegraphics[width=0.45\textwidth]{figs/sinr_ecc.pdf}
% \caption{\small Comparison of SINR values generated by selected path loss models with values collected from OTA data (black). 
% }
% % \vspace{-4mm}
% \label{fig:results_sinr_ecc}
% \end{center}
% \end{figure}

% \begin{figure}[t]
% \begin{center}
% \includegraphics[width=0.45\textwidth]{figs/sinr_ericsson.pdf}
% \caption{\small Comparison of SINR values generated by selected path loss models with values collected from OTA data (black). 
% }
% % \vspace{-4mm}
% \label{fig:results_sinr_ericsson}
% \end{center}
% \end{figure}

The performance of models trained to predict optimal Gold code length (i.e. Task 2) are shown in Fig.~\ref{fig:gold_pred_accuracy} as a series of confusion matrices. For each plot, the optimal Gold code lengths estimated by each model are shown on the x-axis, and the ground truth labels are shown in the y-axis. The decimal ratio of correct predictions for each label are listed on the tile corresponding to that combination of predicted and ground truth labels. 
The performance of the ``real-to-real" model (ESN-2-OTA) is shown in Fig.~\ref{fig:gold_pred_accuracy}(a), which demonstrates perfect prediction accuracy for code lengths of 2047, 511, and 127, and very good accuracy for code lengths of 63 and 31. 
The performance of ESN-2-Friis is shown in Fig.~\ref{fig:gold_pred_accuracy}(b), which shows very good prediction accuracy for length 2047 but poor accuracy for every other length. 
The results for ESN-2-ECC, shown in Fig.~\ref{fig:gold_pred_accuracy}(c), demonstrate poor prediction accuracy for all labels, and indicate nearly random prediction of optimal code length. 
The ESN-2-Ericsson model performance in Fig.~\ref{fig:gold_pred_accuracy}(d) is similar to that of the ESN-2-Friis model, in that both models predict optimal gold code length of 2047. While the selection of this code length will likely provide the best SINR, as discussed in Sec.~\ref{sec:design}, this may not be the optimal choice in scenarios in which link quality is sufficient to support higher data rates.

\section{Discussions}\label{sec:discussion}

% While the POWDER platform provides significant capability for research into protocol design and implementation, network configuration and optimization, and spectrum sharing, we have encountered several key limitations, outlined below. 
In this section, we summarize the key insights gained from our measurements on the POWDER platform as follows.
% \begin{itemize}

    $\bullet$ \textit{Spectrum Priority}. 
    % In the CBRS band, POWDER possesses the lowest-priority spectrum access license, which means users can transmit at a fixed power level, and any incumbent users have resource priority, which allows for significant interference during scheduled experiments. We found that we frequently needed to repeat experiments over several days, leading to scheduling conflicts with other users, to generate sufficient data for both tasks. Furthermore, even using spread spectrum techniques, the low transmit power limited the communication range of the rooftop nodes significantly. \textit{We recommend a future DT framework to include automatic spectrum sharing to develop recommendations for opportunistic resource periods based on historical data, as well as a high-fidelity virtual environment to enable offline experimentation when the RF environment is occupied}.
    In the CBRS band, POWDER operates with the lowest-priority access, meaning incumbent users can cause significant interference during scheduled experiments. As a result, we had to repeat experiments over several days, leading to scheduling conflicts. The low transmit power also limited the communication range of the rooftop nodes. For future DT frameworks, we recommend incorporating automatic spectrum sharing based on historical data and a high-fidelity virtual environment to support offline experimentation when the RF environment is in use.
    
    $\bullet$ \textit{Hardware Configuration Monitoring}. 
    % During data collection, we found that one of the antenna assignments in Shout did not match the current the hardware configuration. While we were able to detect the issue after analyzing the collected data and had time to conduct manual Shout measurements for this node, this inconsistency could be more significant for users implementing their own experimental profiles. \textit{We recommend a future DT system to include active monitoring and real-time statues of all hardware components to automatically identify such inconsistencies}. 
    During data collection, we discovered a mismatch between the antenna assignment in Shout and the current hardware configuration. While we were able to identify and correct this issue after analyzing the data and conducting manual measurements, such inconsistencies could pose a greater challenge for users running custom experimental profiles. We recommend that future DT systems include active monitoring and real-time status updates of all hardware components to automatically detect and resolve such discrepancies.
    
    $\bullet$ \textit{Channel Sounding}. 
    % We found that the Shout framework, while fully open-source, offered poor documentation to support extensive reconfiguration of the protocol. While there is ample documentation on the use of the Shout profiles for general channel measurement, there is very little documentation regarding the role of specific parameters, requiring extensive trial-and-error analysis to reach complete understanding. \textit{We recommend extensive documentation of all user-controllable parameters of both physical and virtual entities in future wireless network DT systems to streamline the generation of OTA data for sim-to-real validation of performance under a broad range of configurations}. 
    We found the Shout framework’s documentation insufficient for extensive reconfiguration, leading to a time-consuming trial-and-error process. While general usage is well-documented, specifics on parameters are lacking. We recommend comprehensive documentation of all user-controllable parameters in future DT systems to facilitate efficient OTA data generation.

\section{Conclusions}\label{sec:conclusion}

% In this work, we have investigated the sim-to-real gap associated with developing a digital twin for the NSF PAWR Platform, POWDER. We first developed a 3D model of the University of Utah campus, incorporating geographical measurements and all rooftop POWDER nodes. We then assessed the accuracy of various path loss models used in training modeling and control policies, examining the impact of each model on sim-to-real link performance predictions. Lastly, we have discussed the lessons learned from model selection and simulation design, and made several recommendations for the future design and implementation of DT-enabled wireless networks. 
% %
% The provided evaluation of the sim-to-real gap can be used to improve the experimental workflow associated with the POWDER testbed, and serve as a roadmap for the continued design and development of DT frameworks. In future work, we will continue evaluation of the sim-to-real gap associated with the POWDER testbed in different scenarios, considering more node types such as mobile nodes and ground-level endpoints, as well as expanding the investigation to include more simulation-based modeling techniques such as ray tracing. 

In this work, we investigated the sim-to-real gap in developing a digital twin for the NSF PAWR Platform, POWDER. We created a 3D model of the University of Utah campus, including geographical measurements and rooftop POWDER nodes, and assessed various path loss models for training and control policies. Our findings highlight the impact of model selection on sim-to-real link performance predictions and provide recommendations for improving DT-enabled wireless network design.
% Our evaluation offers insights for enhancing the POWDER testbed experimental workflow and serves as a guide for future DT framework development. 
In future work we will expand on this work by exploring additional node types, such as mobile nodes and ground-level endpoints, and incorporating advanced simulation techniques like ray tracing.

\bibliographystyle{ieeetr}

\bibliography{biblio}

\begin{thebibliography}{10}
\providecommand{\url}[1]{#1}
\csname url@samestyle\endcsname
\providecommand{\newblock}{\relax}
\providecommand{\bibinfo}[2]{#2}
\providecommand{\BIBentrySTDinterwordspacing}{\spaceskip=0pt\relax}
\providecommand{\BIBentryALTinterwordstretchfactor}{4}
\providecommand{\BIBentryALTinterwordspacing}{\spaceskip=\fontdimen2\font plus
\BIBentryALTinterwordstretchfactor\fontdimen3\font minus
  \fontdimen4\font\relax}
\providecommand{\BIBforeignlanguage}[2]{{%
\expandafter\ifx\csname l@#1\endcsname\relax
\typeout{** WARNING: IEEEtran.bst: No hyphenation pattern has been}%
\typeout{** loaded for the language `#1'. Using the pattern for}%
\typeout{** the default language instead.}%
\else
\language=\csname l@#1\endcsname
\fi
#2}}
\providecommand{\BIBdecl}{\relax}
\BIBdecl

\bibitem{mcmanus2023survey}
M.~McManus, Y.~Cui, J.~Hu, S.~K. Moorthy, Z.~Guan, N.~Mastronarde, E.~S.
  Bentley, and M.~Medley, ``{Digital Twin-Enabled Domain Adaptation for
  Zero-Touch {UAV} Networks: Survey and Challenges},'' \emph{Computer
  Networks}, vol. 236, p. 110000, Nov. 2023.

\bibitem{WenSunReducing20}
W.~Sun, H.~Zhang, R.~Wang, and Y.~Zhang, ``{Reducing Offloading Latency for
  Digital Twin Edge Networks in 6G},'' \emph{IEEE Trans. on Vehicular
  Technology}, vol.~69, no.~10, pp. 12\,240--12\,251, Oct. 2020.

\bibitem{LeiLei21uav}
L.~Lei, G.~Shen, L.~Zhang, and Z.~Li, ``{Toward Intelligent Cooperation of
  {UAV} Swarms: When Machine Learning Meets Digital Twin},'' \emph{IEEE
  Network}, vol.~35, no.~1, pp. 386--392, Jan./Feb. 2021.

\bibitem{Haozhe20}
H.~Wang, Y.~Wu, G.~Min, and W.~Miao, ``{A Graph Neural Network-based Digital
  Twin for Network Slicing Management},'' \emph{IEEE Trans. on Industrial
  Informatics}, vol.~18, no.~2, pp. 1367--1376, Dec. 2020.

\bibitem{NEXTCOMCOM}
J.~Hu, Z.~Zhao, M.~McManus, S.~K. Moorthy, Y.~Cui, N.~Mastronarde, E.~S.
  Bentley, M.~Medley, and Z.~Guan, ``{NeXT: Architecture, Prototyping and
  Measurement of a Software-Defined Testing Framework for Integrated RF Network
  Simulation, Experimentation and Optimization},'' \emph{Journal of Computer
  Communications}, Oct. 2023.

\bibitem{testolina2024bostonTwin}
P.~Testolina, M.~Polese, P.~Johari, and T.~Melodia, ``{Boston Twin: the Boston
  Digital Twin for Ray-Tracing in 6G Networks},'' in \emph{ACM Multimedia
  Systems Conference}, Bari, Italy, Apr. 2024.

\bibitem{villa2024colosseumDT}
D.~Villa, M.~Tehrani-Moayyed, C.~Robinson, L.~Bonati, P.~Johari, M.~Polese, and
  T.~Melodia, ``{Colosseum as a Digital Twin: Bridging Real-World
  Experimentation and Wireless Network Emulation},'' \emph{IEEE Trans. on
  Mobile Computing (Early Access)}, pp. 1--17, Jan. 2024.

\bibitem{bertizzolo2020arena}
L.~Bertizzolo, L.~Bonati, E.~Demirors, A.~Al-shawabka, S.~D'Oro, F.~Restuccia,
  and T.~Melodia, ``{Arena: A 64-antenna {SDR}-based Ceiling Grid Testing
  Platform for Sub-6 GHz 5G-and-Beyond Radio Spectrum Research},''
  \emph{Computer Networks}, vol. 181, p. 107436, Nov. 2020.

\bibitem{pinyo2021sim2realtransfer}
P.~Pinyoanuntapong, T.~Pothuneedi, R.~Balakrishnan, M.~Lee, C.~Chen, and
  P.~Wang, ``{Sim-to-Real Transfer in Multi-agent Reinforcement Networking for
  Federated Edge Computing},'' in \emph{IEEE/ACM Symposium on Edge Computing},
  San Jose, CA, USA, Dec. 2021.

\bibitem{gao2023wiprox}
Y.~Gao, G.~Chi, G.~Zhang, and Z.~Yang, ``{Wi-Prox: Proximity Estimation of
  Non-Directly Connected Devices via Sim2Real Transfer Learning},'' in
  \emph{IEEE Global Communications Conference}, Kuala Lumpur, Malaysia, Dec.
  2023.

\bibitem{breen2020powder}
J.~Breen \emph{et~al.}, ``{Powder: Platform for Open Wireless Data-driven
  Experimental Research},'' in \emph{Proc. of International Workshop on
  Wireless Network Testbeds, Experimental evaluation and Characterization
  (WiNTECH}, London, United Kingdom, Sept. 2020.

\bibitem{utah_gis_map}
\BIBentryALTinterwordspacing
U.~G. Survey, ``National geologic map database project: Fort douglas, ut,''
  Accessed Mar. 8, 2023. 2020. [Online]. Available:
  \url{https://ngmdb.usgs.gov/topoview/viewer/\#14/40.7652/-111.8492}
\BIBentrySTDinterwordspacing

\bibitem{Webb2021Shout}
K.~Webb, S.~K. Kasera, N.~Patwari, and J.~K.~V. der Merwe, ``{{WiMatch:}
  Wireless Resource Matchmaking},'' in \emph{IEEE Conference on Computer
  Communications Workshops}, Vancouver, BC, Canada, May 2021.

\bibitem{bianchi2021esn}
F.~M. Bianchi, S.~Scardapane, S.~L{\o}ske, and R.~Jenssen, ``{Reservoir
  Computing Approaches for Representation and Classification of Multivariate
  Time Series},'' \emph{IEEE Trans. on Neural Networks and Learning Systems},
  vol.~32, no.~5, pp. 2169 -- 2179, June 2021.

\bibitem{pyESNgithub}
\BIBentryALTinterwordspacing
{C. Kornd{\"o}rfer}. {{pyESN}: Echo State Networks in Python}. [Online].
  Available: \url{https://github.com/cknd/pyESN}
\BIBentrySTDinterwordspacing

\end{thebibliography}

\end{document}